\documentclass[twocolumn]{article}
\usepackage{amsmath}
\usepackage{epsfig}

\begin{document}

\noindent
{\large\bf Comment on ``Quantum Games and Quantum Strategies''}
\smallskip

\begin{figure}[t]
\includegraphics[width=.48\textwidth]{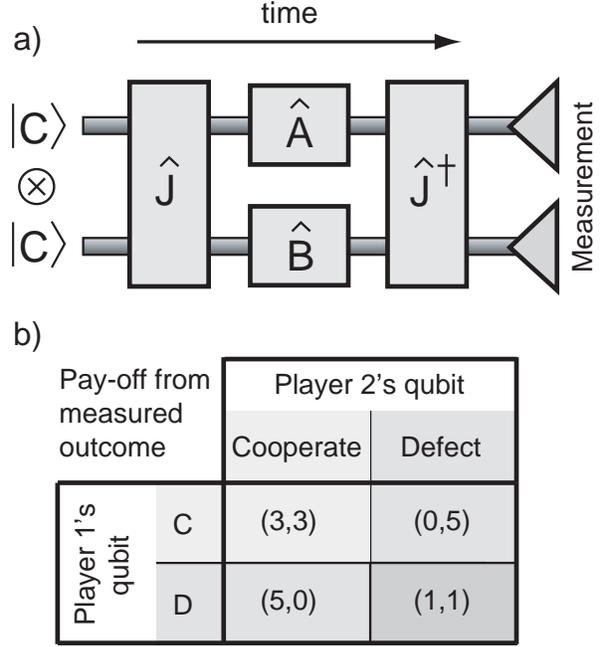}
\caption{a) The quantized Prisoner's Dilemma, as described in
\cite{Eisert}.  The pair of qubits are prepared in the unentangled 
state $\left|CC\right\rangle$ and then sent through the entangling gate 
$\hat{J}$.  Players $A$ and $B$ then apply their local unitary operations 
$\hat{A}\otimes\hat{I}$ and $\hat{I}\otimes \hat{B}$, respectively.
A gate inverse to $\hat{J}$ is applied before the final measurement.
b) The Prisoner's Dilemma pay-off table chosen in \cite{Eisert}.}
\label{figure1}
\end{figure}

In a recent Letter, Eisert {\em et al.} \cite{Eisert} presented a quantum mechanical
generalization of Prisoner's Dilemma. In the classical form of this game, rational analysis
leads the two players to `defect' against one-another in a mutually destructive fashion
\cite{classicalPDref}. A central result of Eisert {\em et al.}'s Letter is the
observation that their quantum variant, illustrated in Figure \ref{figure1}, 
can avoid the `dilemma':
the mutually destructive outcome is replaced with an effectively cooperative one. 
Specifically, it is asserted that the maximally entangled game's 
unique Nash equilibrium \cite{classicalPDref}
occurs when both players apply the strategy $\hat{Q} = i \sigma_z$,
yielding a pay-off
equivalent to cooperative behaviour in the classical game.

In this Comment we show that their observation is incorrect. The mistake
follows from the following erroneous assertion:

\begin{quotation}
It proves to be sufficient to restrict the strategic space to the 
2-parameter set of unitary 2 x 2 matrices,
\begin{equation}
\hat{U}(\theta,\phi) = \left( \begin{array}{cc}
e^{i\phi} {\rm cos} \theta/2 & {\rm sin} \theta/2 \\ 
-{\rm sin} \theta/2 & e^{-i\phi} {\rm cos} \theta/2
\end{array} \right),
\end{equation}
with $0 \leq \theta \leq \pi$ and $0 \leq \phi \leq \pi / 2$.
\end{quotation}

Here we explicitly consider the
complete set of \emph{all} local unitary operations (\emph{i.e.} all of
$SU(2)$), finding that the properties of the game are wholly different: the strategy
$\hat{Q}$ is not an equilibrium; indeed, there is {\em no equilibrium} in the space of
deterministic quantum strategies. 
Moreover, it seems unlikely that the restriction to the set
$\hat{U}(\theta,\phi)$ can reflect any reasonable physical constraint (limited
experimental resources, say) because this set is not closed under 
composition.  An ideal counter strategy to $\hat{Q}$, for example, is 
$i\sigma_x$, which is equal to $\hat{U}(0,\pi/2)\hat{U}(\pi,0)$.
The game of \cite{Eisert} therefore does not constitute a reasonable 
variant of the general case we consider here.

We will write the operations applied by the players in the form 
$\hat{A} \otimes \hat{B}$, where $\hat{A}$ is applied to the qubit controlled 
by $A$ and similarly for $\hat{B}$.
Suppose that player $A$
applies transformation
$\hat{X}$ to her qubit, prepared as the first qubit in the maximally 
entangled state
\begin{equation}
\hat{J} \left| CC\right\rangle = \frac{1}{\sqrt{2}}( \left| CC\right\rangle + i \left| DD\right\rangle ),
\end{equation}
where 
$\hat{J}=\exp\{i\pi\hat{D}\otimes\hat{D}/4\}$ and $\hat{D}=i\sigma_y$ is 
the `defect' matrix of \cite{Eisert}. The most general 
$\hat{X} \in SU(2)$ is of the form
$\hat{X}=(x_{ij})$, where $x_{11}=x_{22}^*$, $x_{12} = -x_{21}^*$ and
$\det{\hat{X}}=1$.  
Therefore,
$A$ produces the state $(\hat{X} \otimes \hat{I})\hat{J}\left| CC \right\rangle = (I \otimes \hat{Y})J\left| CC \right\rangle$ 
for $\hat{Y} = (y_{ij}) \in SU(2)$, where $y_{11} = x_{11}$ and $y_{12} = i x_{12}$.  In other words, any unitary
transformation  which $A$ applies locally to her qubit is actually equivalent to a unitary transformation
applied locally by $B$.  Consequently, if  $B$ were to choose
$\hat{D}\hat{Y}^\dagger$, we would have a final state
$\hat{J}^\dagger(\hat{X}\otimes \hat{D}\hat{Y}^\dagger)\hat{J}\left| CC \right\rangle
=\hat{J}^\dagger(\hat{I}\otimes \hat{D}\hat{Y}^\dagger\hat{Y})\hat{J}\left| CC \right\rangle
= \left| CD \right\rangle$, the optimal outcome for $B$. Thus, for any given strategy of $A$,
there is an ideal counter-strategy for $B$, and vice-versa; there is no Nash equilibrium of the kind
suggested by Eisert {\em et al.} \cite{ftNote1}.

To obtain such equilibria we must extend the
abilities of the players: it suffices to allow them to make 
probabilistic choices (rather than the full formalism of completely 
positive maps
considered in footnote 14 of \cite{Eisert}).
Suppose that
$A$ adopts the following strategy: she will choose a 
move $\hat{X} \in SU(2)$ at random with respect to the 
Haar measure on $SU(2)$.  
If $B$ responds with $\hat{Y}_0 \in SU(2)$ then the probability that 
outcome 
$i \in \{CC,CD,DC,DD\}$ will be measured is
\begin{align}
P_i(\hat{Y}_0) 
&= \int_{SU(2)} \left| 
	\left\langle i \right|\hat{J}^\dagger (\hat{X} \otimes \hat{Y}_0) 
	\hat{J} \left| CC \right\rangle \right|^2 d\hat{X} \nonumber \\
&= \int_{SU(2)} \left|
	\left\langle i \right|\hat{J}^\dagger (\hat{X} \hat{X}_0^\dagger \hat{X}_0 \otimes I)
	\hat{J} \left| CC \right\rangle \right|^2 d\hat{X} \nonumber \\
&= \int_{SU(2)} \left|
	\left\langle i \right|\hat{J}^\dagger (\hat{X} \otimes \hat{I}) 
	\hat{J} \left| CC \right\rangle \right|^2 d\hat{X} \nonumber \\
&= P_i(\hat{I}),
\end{align}
where $\hat{X}_0 \in SU(2)$ is chosen such that 
$(\hat{X}_0 \otimes \hat{I}) \hat{J} \left| CC \right\rangle = 
(\hat{I}\otimes\hat{Y}_0) \hat{J} \left| CC \right\rangle$ and
we have used the right invariance of the Haar measure, assumed to
be normalized such that the volume of $SU(2)$ is $1$.  Thus, 
$B$'s choice of strategy does not matter; regardless of his choice,
his expected pay-off is simply an unbiased average over the classical pay-offs.
Therefore, the situation where both players adopt this random strategy
is a Nash equilibrium: neither player can improve his or her payoff by
unilaterally altering choice of strategy.

As a final point, we note that one can construct 
Prisoner's Dilemma-type pay-off tables
for which the quantum equilibrium pay-off we describe above is below the 
classical equilibrium pay-off, or above it, or even above the classical
cooperative pay-off. In this last case the `dilemma' may be 
said to
have been removed~\cite{ftNote2}. 
To this extent the behavior of the quantum generalization
is qualitatively different from the classical case, although in a way that 
is perhaps less surprising than originally suggested by Eisert \emph{et al.}

We thank Neil Johnson for helpful conversations.  SB is supported by EPSRC. PH
acknowledges the support of the Rhodes Trust.

\bigskip
\noindent{\small Simon C. Benjamin and Patrick M. Hayden}

{\small Centre for Quantum Computation}

{\small University of Oxford}

{\small Clarendon Laboratory, Parks Road }

{\small Oxford OX1 3PU, UK}



\begin{thebibliography}{30}
\bibitem{Eisert}  J. Eisert, M. Wilkens, M. Lewenstein, 
\emph{Phys. Rev. Lett.} {\bf 83} 3077 (1999), quant-ph/9806088. 
\bibitem{classicalPDref} R. B. Myerson, 
\emph{Game Theory: An Analysis of Conflict} (MIT Press, Cambridge, MA, 1991).
\bibitem{ftNote1}
This result is a familiar property of maximally entangled states
and is easily generalized to two maximally
entangled $n$-dimensional systems $\cal{H}_A$ and $\cal{H}_B$.  
Such states can always be written in the form
\begin{equation}
\left|\Psi\right\rangle = \frac{1}{\sqrt{n}} \sum_{i=1}^n \left| i 
\right\rangle_A \otimes \left| i \right\rangle_B,
\end{equation}
where $_A\left\langle i | j \right\rangle_A = {_B\left\langle i | j \right\rangle_B} =
\delta_{ij}$.
It is then easy to check that if $U \in SU(\cal{H}_A)$,
$U \otimes I \left|\Psi\right\rangle = I \otimes U^T \left|\Psi\right\rangle$.
Since the special unitary group
is closed under transpose, it follows that our
observations on the present quantum game, the classical form of which 
involves the one-bit strategies $\{C,D\}$, will also apply to 
quantum generalizations of any larger classical two-player
symmetric games.
\bibitem{ftNote2} For the pay-off values (3,5,0,1) given in \cite{Eisert}, 
the quantum
equilibrium pay-off (QP) is 2.25, in-between the classical equilibrium pay-off
(CEP) of 1 and the classical cooperative pay-off (CCP) of 3. However, for the
values (5,6,0,4) the QEP is below the CEP, whilst for (3,9,0,1) it is above the
CCP.
\end{thebibliography}
\end{document}